\documentclass{jfm}

\usepackage{graphicx}
\usepackage{natbib}
\usepackage{amsmath}

\ifCUPmtlplainloaded \else
  \checkfont{eurm10}
  \iffontfound
    \IfFileExists{upmath.sty}
      {\typeout{^^JFound AMS Euler Roman fonts on the system,
                   using the 'upmath' package.^^J}%
       \usepackage{upmath}}
      {\typeout{^^JFound AMS Euler Roman fonts on the system, but you
                   dont seem to have the}%
       \typeout{'upmath' package installed. JFM.cls can take advantage
                 of these fonts,^^Jif you use 'upmath' package.^^J}%
      }
  \else
  \fi
\fi

\ifCUPmtlplainloaded \else
  \checkfont{msam10}
  \iffontfound
    \IfFileExists{amssymb.sty}
      {\typeout{^^JFound AMS Symbol fonts on the system, using the
                'amssymb' package.^^J}%
       \usepackage{amssymb}%

      }{}
  \fi
\fi

\ifCUPmtlplainloaded \else
  \IfFileExists{amsbsy.sty}
    {\typeout{^^JFound the 'amsbsy' package on the system, using it.^^J}%
     \usepackage{amsbsy}}
    {}
\fi

\newsavebox{\astrutbox}
\sbox{\astrutbox}{\rule[-5pt]{0pt}{20pt}}

\def\der#1#2{{\partial #1\over \partial #2}}

\def\be{\begin{equation}}
\def\ee{\end{equation}}
\def\bea{\begin{eqnarray}}
\def\eea{\end{eqnarray}}
\def\bse{\begin{subequations}}
\def\ese{\end{subequations}}
\def\bsea{\begin{subeqnarray}}
\def\esea{\end{subeqnarray}}
\def\({\left (}
\def\){\right )}
\def\[{\left [}
\def\]{\right ]}
\def\<{\left <}
\def\>{\right >}

\title[On the role of vortex stretching in optimal growth]{On the role of vortex stretching in energy optimal growth of three dimensional perturbations on plane parallel shear flows}

\author[H. Vitoshkin, E. Heifetz,  A. Yu. Gelfgat and N. Harnik]%
{H.\ns V\ls I\ls T\ls O\ls S\ls H\ls K\ls I\ls N$^1$ E.\ns H\ls E\ls I\ls F\ls E\ls T\ls Z$^2$%
  \thanks{Email address for correspondence: eyalh@post.tau.ac.il},\ns
A.\ns Yu.\ns G\ls E\ls L\ls F\ls G\ls A\ls T$^1$\break\texttt{}
\and N.\ns H\ls A\ls R\ls N\ls I\ls K$^2$}

\affiliation{$^1$School of Mechanical Engineering, Faculty of Engineering, Tel Aviv University, Israel\\[\affilskip]
$^2$Department of Geophysics and Planetary Sciences, Tel-Aviv University, Israel}

\pubyear{2010}
\volume{650}
\pagerange{119--126}
\date{?; revised ?; accepted ?. - To be entered by editorial office}
\begin{document}

\maketitle

\begin{abstract}

The three dimensional optimal energy growth mechanism, in plane parallel shear flows, is reexamined in terms of the role of vortex stretching and the interplay between the span-wise vorticity and the planar divergent components. For high Reynolds numbers the structure of the optimal perturbations in Couette, Poiseuille, and mixing layer shear profiles is robust and resembles localized plane-waves in regions where the background shear is large. The waves are tilted with the shear when the span-wise vorticity and the planar divergence fields are in (out of) phase when the background shear is positive (negative). A minimal model is derived to explain how this configuration enables simultaneous growth of the two fields, and how this mutual amplification reflects on the optimal energy growth. This perspective provides an understanding of the three dimensional growth solely from the two dimensional dynamics on the shear plane.

\end{abstract}


\section{Introduction}
It is well recognized that three dimensional (3D) perturbations at incompressible, high Reynolds number, on plane parallel shear flows, attain non-modal growth which may be much larger than attained by two dimensional (2D) perturbations that are confined to the shear plane, e.g. \citet{Butler92} for Couette and \citet{Reddy93} for Poiseuille shear flows. The 3D growth appears at later stages when the perturbations are tilted with the background shear. This stands in contrast with the 2D optimal growth which is obtained when the perturbations are tilted against the shear. In 2D, the perturbation in the energy norm grows via the Orr mechanism \citet {Orr07} and, in the presence of an inflection point, by the action at a distance between counter-propagating Rossby waves \citet{Heifetz2005}. The 3D growth mechanism is commonly rationalized by the lift-up mechanism \citet{Landhall80}, \citet{SchmidandHenningson2001}, which can also be viewed as a tilt-up of the span-wise background vorticity by the cross-stream perturbation velocity \citet{Farrell93}. This explanation follows the mathematical procedure by which the dynamics is usually resolved -- a homogeneous equation for the cross-stream velocity is derived, and the span-wise variation of this velocity serves as a source for the tilt-up of the background vorticity.

However, the third dimension adds a fundamental mechanisms which is absent from strictly 2D flow -- the background shear vorticity (pointing by definition to the span-wise direction) may be stretched due to contraction of areas in the shear plane by the planar divergence of shear plane projection of the perturbation field ($d$). This implies, by conservation of circulation, a generation of span-wise component of vorticity perturbation ($q$). Thus, the perturbation divergence and vorticity scalar fields evolve together, and we expect optimal growth to occur when the interplay between $d$ and $q$ results in a simultaneous growth of the two fields. Since the circulation associated with $q$ is on the shear plane, the interplay between $d$ and $q$ is expected to provide an understanding of the 3D optimal growth solely in terms of the 2D planar perturbation dynamics.

In section 2 we show that it is a robust feature that the largest 3D growth is obtained when $d$ and $q$ are in (anti) phase in regions of positive (negative) mean shear. Furthermore, the fastest growing perturbations resemble localized plane-waves that are tilted with the local maximal shear and this behavior is insensitive to the shear curvature. In section 3 we therefore derive a minimal model for the interplay between $d$ and $q$ in the presence of a constant background shear, for plane-waves. In Section 4 the optimal growth in the energy norm is analyzed from this $d$ -- $q$ perspective, and in Section 5 we conclude our results.

\section{Numerical comparison between 2D and 3D growths}\label{sec:2D and 3D}

We define the cartesian coordinates ${\bf r} =  (x{\bf i},y{\bf j},z{\bf k})$ as the (stream-wise, cross-stream, span-wise) directions so that the background shear velocity is $\overline{\bf U} = \overline{U}(y){\bf i}$, the perturbation velocity vector is ${\bf u}  = (u,v,w)$, the perturbation span-wise vorticity is $q = \der{v}{x} - \der{u}{y}$, and the 2D divergence field on the shear plane is $d=  \der{u}{x} + \der{v}{y}$.

In Fig. 1. we present the optimal evolutions of 2D and 3D perturbations in plane parallel bounded Couette and Poiseuille shear flows, as well as in an open mixing layer profile. The first two examples are in excellent agreement with the results of \citet{Reddy93}, whereas the calculations of the latter are new (to the best of our knowledge), and appear in more detail in a companion paper by \citet{Vitoshkin2012}. The calculations are done for relatively high Reynolds numbers, but below the critical values enabling modal instability. The optimal vectors of the perturbations for Fourier modes in the stream-wise and span-wise directions ($\propto e^{i(kx +mz)}$) are computed, where in the cross-stream direction the perturbations are discretised and resolved by standard finite difference schemes. Indeed, for the three profiles the 3D maximal growths are larger by an order of magnitude than the correspondent 2D ones, and are attained at later stages. The $q$ contours indicate that in all cases the eddies are initially tilted against the shear, and then they evolve to become more aligned.

The major difference between the growth evolution of 2D and 3D flows, is that in the former the maximum energy value is obtained when the eddies are aligned perpendicular to the shear, whereas in the latter it is obtained when the eddies are tilted with the shear. For the 3D perturbations, the contours of $d$ are superposed (for 2D perturbations $d$ is zero since incompressibility is assumed). It is evident for the three profiles, that when the eddies experience their largest 3D growth, $d$ and $q$ are in (anti) phase when the mean shear is positive (negative), and the perturbation structures resemble localized plane-waves that are tilted with the local maximal shear.
These structures are robust and were found in many different Fourier modes of the three canonical profiles. Therefore, it seems that at the stage of optimal growth the exact curvature of the shear profile is not too important. The essence of the $d$  -- $q$ interaction, is thus examined next in a minimal model of a plane-wave in the presence of a constant background shear.

\begin{figure}
\centering{\includegraphics[scale=0.55]{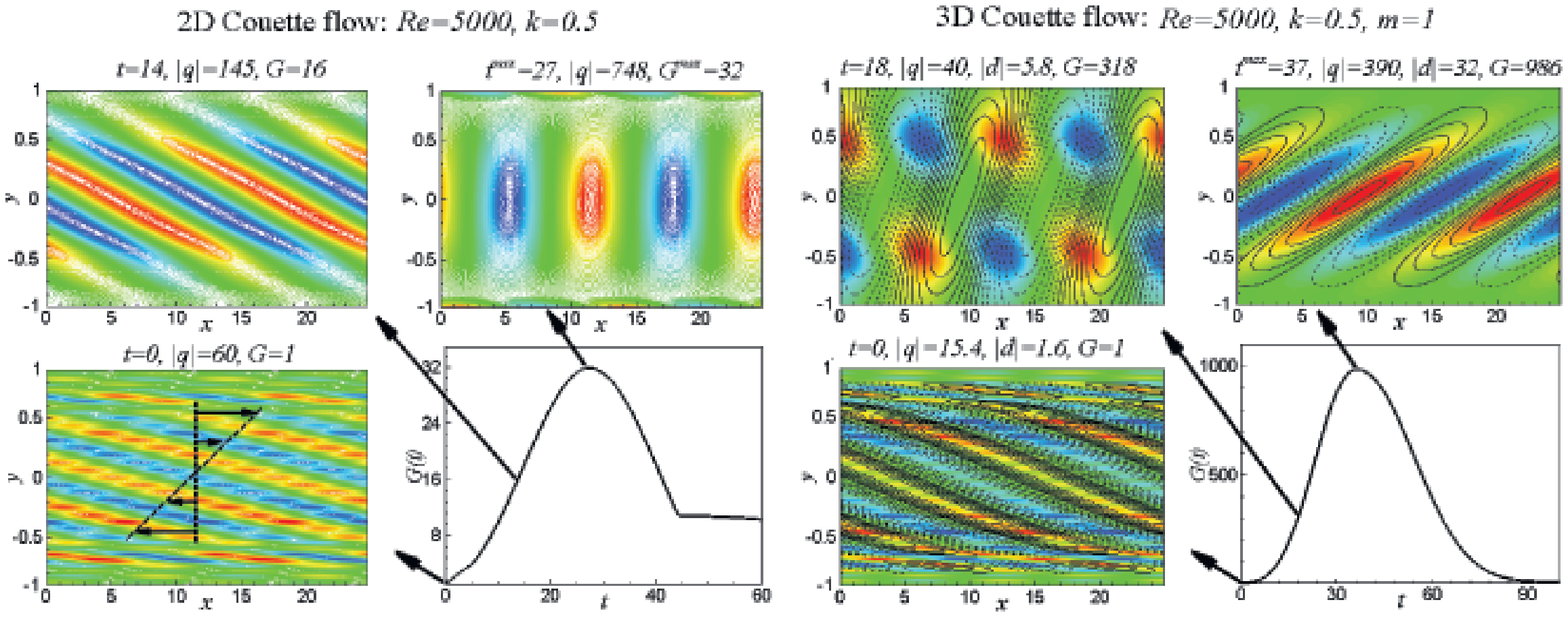}
           \includegraphics[scale=0.55]{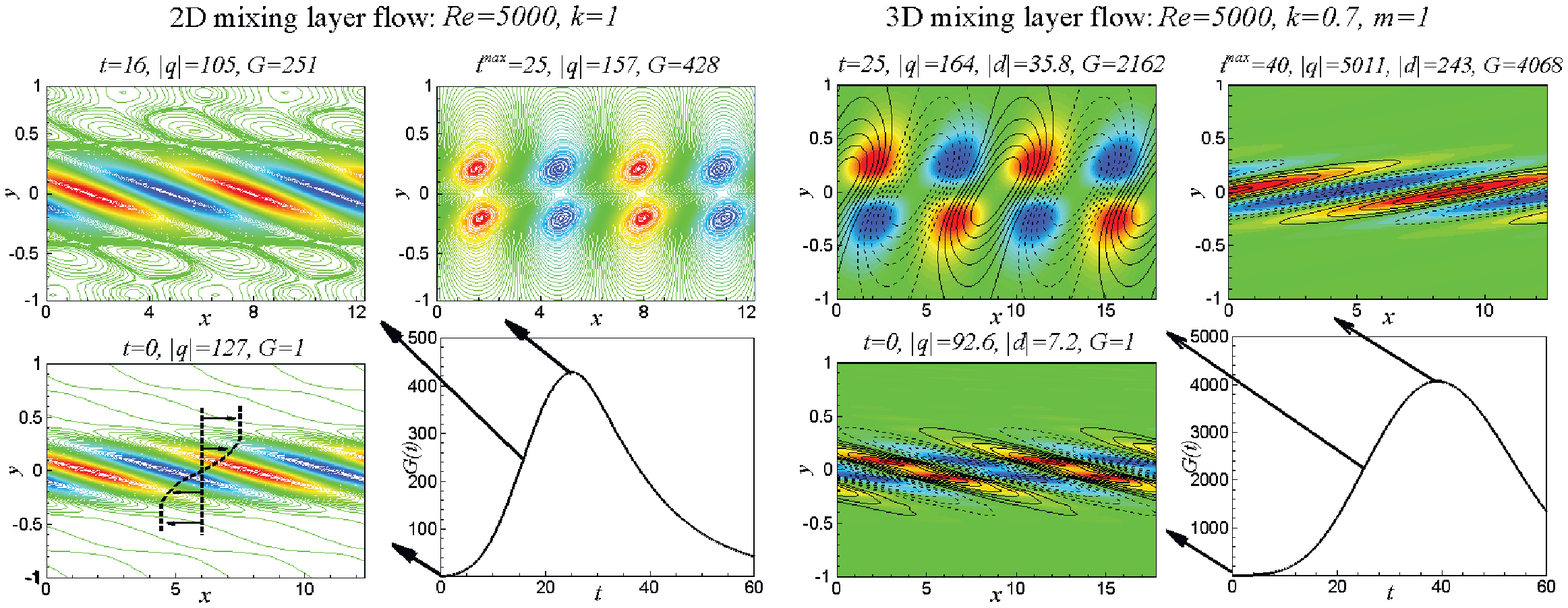}           \includegraphics[scale=0.53]{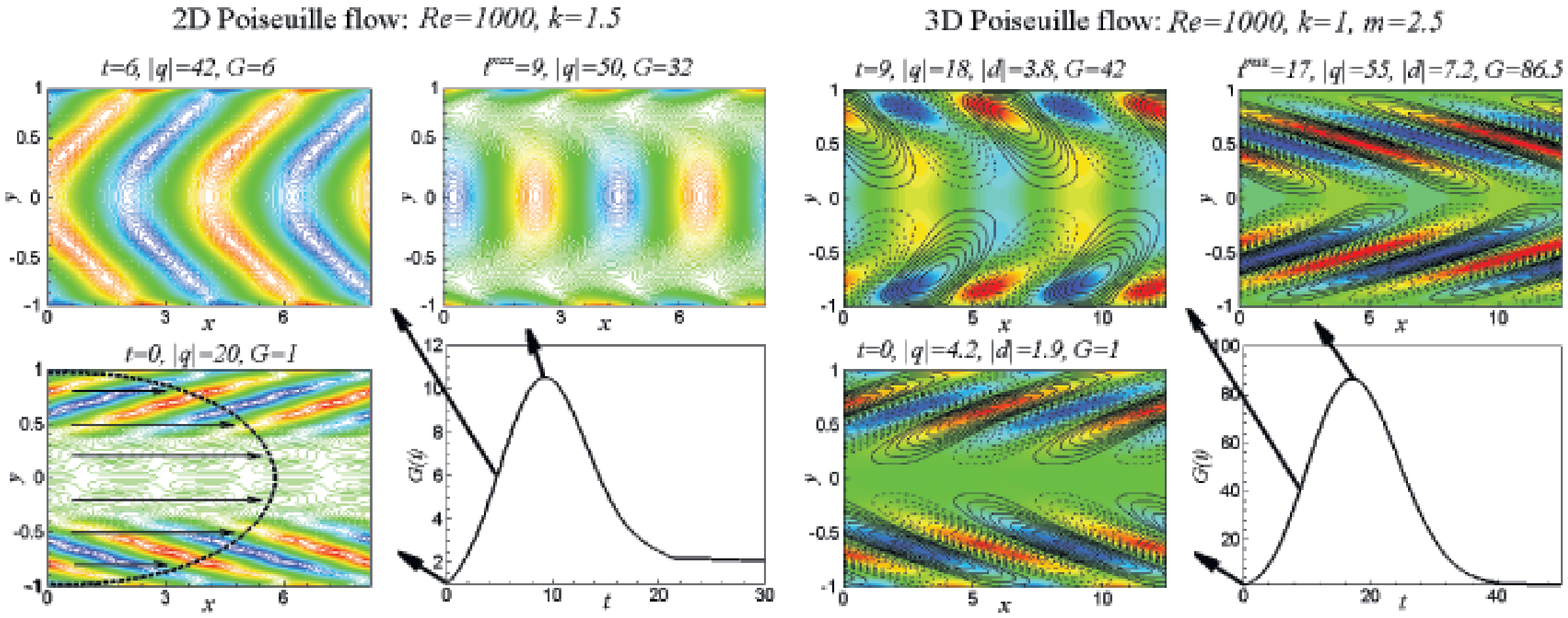}}
\caption{Optimal energy growth for 2D (left columns) and 3D (right columns) perturbations on Couette (upper) Poiseuille (middle) and mixing layer (lower) plane parallel shear flows for relatively high Reynolds numbers ($Re = 5000$, $1000$, $5000$, respectively). The different stream-wise and span-wise wavenumbers $(k,m)$ are selected to generate maximal non-modal growths. Solid curves indicate the energy growth evolution, $G(t)$, from the initial optimal perturbations. Note that ${G_{3D}}_{max} >> {G_{2D}}_{max}$ and ${t_{3D}}_{max} > {t_{2D}}_{max}$. The structure of the optimal perturbations on selected times is indicated by the contours of the span-wise vorticity $q$. For the 3D perturbations the planar divergent field $d$ is superposed and indicated by the dashed contours. Note that in 2D, $G_{max}$ is obtained when the eddies are aligned perpendicular to the shear, whereas in 3D it occurs when the eddies are tilted with the shear and $d$ and $q$ are in (anti) phase when the mean shear is positive (negative). Furthermore, the structures resemble localized plane-waves that are tilted with the local maximal shear.}
\label{fig:figure10}
\end{figure}

\section{{\bf An analytical} model of 3D plane-wave growth in constant shear}\label{sec:Analytic model}

We consider an unbounded Eulerian flow with a constant background shear $\overline{U}(y) = \Lambda y$, where  $\Lambda = \der{\overline{U}}{y} = -\overline{q}$, and $\overline{q}$ is the background vorticity, pointing to the span-wise direction. Since in most regions of the three canonical examples the background shear is positive, $\Lambda$ is determined to be a positive constant and therefore the background span-wise vorticity $\overline{q}$ is negative.
The linearized eddy momentum equation can be written then as:
\be
{D{\bf u}\over Dt} = -\nabla{p} -\Lambda v {\bf i}
\ee
where $p$ is the perturbation pressure divided by the constant density of the flow, and ${D\over Dt} = \der{}{t} + \Lambda y\der{}{x}$ is the linearized Lagrangian derivative.

Conservation of circulation implies that when $d$ is positive the absolute value of the span-wise vorticity decreases. Since $\overline{q}$ is negative, a positive anomaly of $q$ should be generated when $d$ is positive (Fig. 2a):
\be
{Dq \over Dt} =\Lambda d
\ee
Thus, in order to obtain growth in $q$, $q$ and $d$ should be positively correlated. On the other hand, the divergence tendency equation, obtained from equation (3.1), is:
\be
{Dd \over Dt} =  -\nabla^2_2 p - 2\Lambda\der{v}{x}
\ee
where $\nabla^2_2 = \der{}{x^2} + \der{}{y^2}$ is the 2D Laplacian on the shear plane. The first term on the right hand side indicates that $d$ will grow when the perturbation pressure anomaly is positive. The second term shows the contribution of differential advection to $d$. The factor of $2$ stems from the separate contributions of the differential advection of the mean flow by the cross-stream perturbation velocity, and the differential advection of the cross-stream perturbation velocity by the mean flow (Fig. 2b). Incompressibility ($\nabla\cdot{\bf u} = 0$) determines however, that when $d$ increases the pressure anomaly is negative since:
\be
{Dd \over Dt} = - {D\over Dt}\der{w}{z} = \der{^2 p}{z^2}
\ee
This somewhat counter-intuitive behavior is simply because a divergent motion in the shear plane must be accompanied by a shrinking in the span-wise direction. The latter can only result from a negative pressure anomaly. This implies that the two terms on the RHS of (3.3) must be of opposite signs with the latter dominating.

\begin{figure}
\centerline{\includegraphics[scale=0.50]{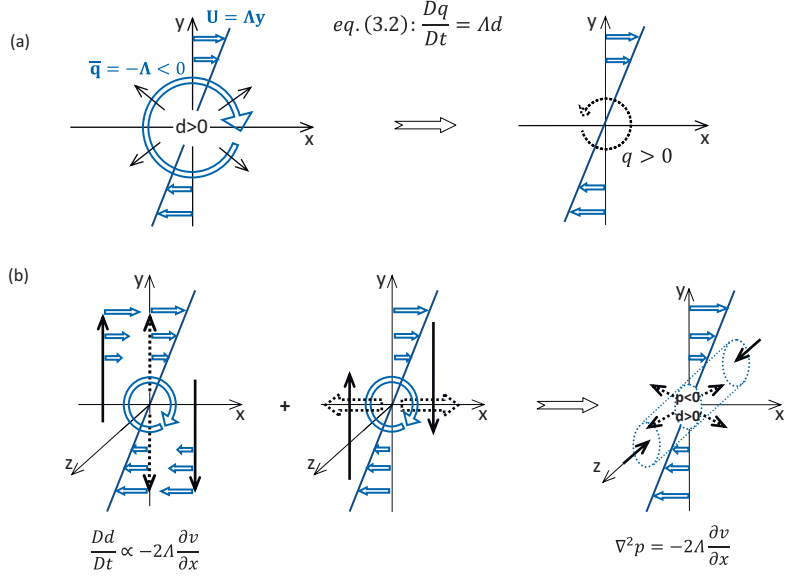}}
\caption{Graphical interpretation of the dynamical processes described in equations (3.2) - (3.4). Doubled and bold single arrows represent the background and the perturbation flows, respectively. Dashed arrows represent the result of the process (a) Equation (3.2): conservation of circulation implies that an area expansion in the shear plane by the planar divergent field perturbation $d$, decreases the span-wise vorticity. Since the background vorticity $\overline{q}$ is negative, expansion generates a positive span-wise vorticity perturbation $q$. (b) The second term on the RHS of (3.3) for $\der{v}{x} < 0$: generation of $d$ by differential advection of the mean flow by the cross-stream perturbation velocity, together with differential advection of the cross-stream perturbation velocity by the mean flow. Equation (3.4) and incompressibility imply that expansion in the shear plane results in a shrinking in the span-wise direction, due to a negative pressure perturbation. Therefore, for a positive background shear $sign{(p)} = sign{(\der{v}{x})}$ and the two terms on the RHS of (3.3) are of opposite signs, with  the latter dominating.}
\label{fig:kd}
\end{figure}

For a plane-wave solution, deformed by the constant shear $\Lambda$, the RHS of (3.3) can be expressed in terms of $q$.
Writing generally:  $\chi({\bf r},t) = \hat{\chi}(t)e^{i{\bf K}\cdot{\bf r}}$, with the 3D wavenumber vector ${\bf K} = (k,l = l_0 -k\Lambda t,m)$, then
${D\chi \over Dt} = {d{\hat{\chi}} \over dt}e^{i{\bf K}\cdot{\bf r}}$. Since equations (3.3) and (3.4) yield the diagnostic relation, $\nabla^2 p = - 2\Lambda\der{v}{x}$ we can write:
\be
{Dd \over Dt} = -2\Lambda\({m\over |{\bf K}|}\)^2\der{v}{x}
\ee
It is now left to express $v$ in terms of $q$ and $d$ by applying the 2D Helmholtz decomposition:
\renewcommand{\theequation}{3.\arabic{equation}a,b}
\be
u = u_d + u_q = \der{\phi}{x} - \der{\psi}{y}, \hspace{0.5cm}
v = v_d + v_q = \der{\phi}{y} + \der{\psi}{x}
\ee
so that
\be
d = \nabla^2_2\phi, \hspace{0.5cm} q = \nabla^2_2\psi
\ee
Equations (3.2) and (3.5) now determine the plane-wave $d$ -- $q$ dynamics:
\be
{d{\hat q}\over dt} =\Lambda {\hat d}, \hspace{0.5cm}
{d{\hat d}\over dt} = -2\Lambda\({m\over |{\bf K}|}\)^2{k \over (k^2+l^2)}(l{\hat d} +k{\hat q})
\ee
where the general solution of (3.8) is given, for completeness, in the Appendix.

We look for configurations allowing simultaneous growth in both $q$ and $d$. As pointed out, (3.8a) implies that $q$ and $d$ must be positively correlated. For positive stream-wise wavenumber $k$ (by construction), (3.8b) indicates that $-\der{v_q}{x}$ is always negative for positive $q$, hence while $d$ acts to increase $q$, $q$ acts to decrease $d$. Therefore, only the cross-stream velocity $v_d$, induced by the divergent field itself, can contribute to the divergent growth. Equation (3.8b) and Fig. 3 indicate that this may happen only when  $l$ is negative, that is when the plane-wave is tilted with the shear, i.e., later than $t = l_0/(k\Lambda)$.

The $d$ -- $q$ dynamics are reflected in the energy growth mechanism via the Reynolds stress:
\be
\der{}{t}\<E\> = -\Lambda\<uv\>, \hspace{0.5cm} E = {|{\bf u}|^2 \over 2}
\ee
where $\<\,\,\,\>$ represents spatial averaging, and
$$
-\<uv\> = -\<(u_q +u_d)(v_q +v_d)\> =-\<(u_q v_q + u_d v_d + u_q v_d + u_d v_q )\> =
$$
\renewcommand{\theequation}{3.\arabic{equation}}
\be
{1\over 2(k^2 +l^2)^2}\[kl\hat{q}^2 -kl\hat{d}^2 + l^2\Re{(\hat{q}\hat{d}^*)} - k^2\Re{(\hat{d}\hat{q}^*)} \]
\ee
The first term represents the 2D Orr mechanism which is positive (negative) when $l$ is positive (negative), whereas the second term represents the ability of the divergent field to amplify (decay) itself for negative (positive) $l$. If $q$ and $d$ are in phase, $[\Re{(\hat{q}\hat{d}^*)} = \Re{(\hat{d}\hat{q}^*)} = |\hat{q}||\hat{d}|]$, the third mixed term is always positive but the fourth one is always negative. The signs of the different terms are evident as well from Fig. 3.

\section{Optimal $d$ -- $q$ dynamics in the canonical profiles}\label{sec: Optimal $d$ -- $q$}

The analysis in the previous section suggests that the structure of the optimal evolution in the early and intermediate stages should not take the form of a plane-wave since when $l>0$, the increasing of $q$ by $d$ is accompanied by a decreasing of $d$ by both $q$ and $d$. Indeed, as indicated from Fig. 1, during these stages $d$ and $q$ depart from each other and do not resemble a plane-wave structure. This uneven interaction is expected to generate a much larger growth in $q$ than in $d$, as presented in Fig. 4a. In Fig. 4b the evolution of the normalized inner product of $d$ and $q$ is shown, at levels of strong shear for the three profiles. Initially $d$ and $q$ are in phase, however while $d$ amplifies $q$, both $q$ and $d$ act to decay $d$ so that at some stage $d$ changes sign and $q$ and $d$ become anti-phased, and then $d$ acts to decay $q$. Nonetheless, at the later stage of maximum growth, $q$ and $d$ return to be in phase and a mutual growth is obtained when the plane-wave like structures are tilted with the shear\footnote{For a non-constant shear, equation (3.2) should be modified to ${Dq \over Dt} =\Lambda d + v\der{^2\overline{U}}{y^2}$. The latter additional term is the advection of the mean flow vorticity by the cross-stream velocity perturbation. It is vital for the 2D optimal growth mechanism in the presence of an inflection point, e.g. Heifetz and Methven, 2005, however, in Fig. 4c we can see that this term contributes almost nothing to the 3D growth mechanism.}.

\begin{figure}
\centering{\includegraphics[scale=0.50]{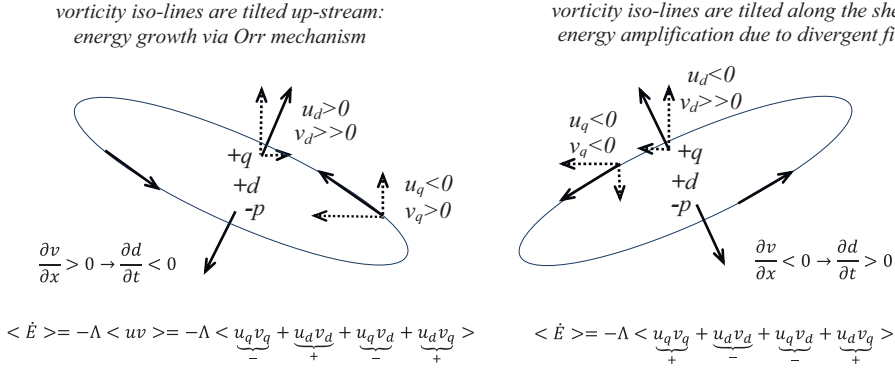}}
\caption{Planar divergent and rotational contributions to the 3D energy growth via the Reynolds stress mechanism, for eddies with positive span-wise vorticity, positive planar divergence and negative pressure perturbations. The contributions are illustrated for both cases where the eddies are tilted negatively against the shear and positively with the shear.
The rotational Orr term, $-<u_qv_q>$, is positive (negative) for negative (positive) tilt, and the divergent term, $-<u_dv_d>$, is positive (negative) for positive (negative) tilt. The sign of the first mixed $d$--$q$ term, $-<u_dv_q>$, behaves as the Orr term, whereas the sign of the second term $-<u_qv_d>$, as the divergent one. When the eddies are strongly tilted with the shear the latter term dominates and generates large growth that overwhelms the strong decay by the Orr mechanism. Furthermore, when the eddies are strongly tilted, $\der{v}{x}<0$,  and simultaneous growths for both $d$ and $q$ are obtained.}
\label{fig:figure7}
\end{figure}

\begin{figure}
\centering{\includegraphics[scale=0.19]{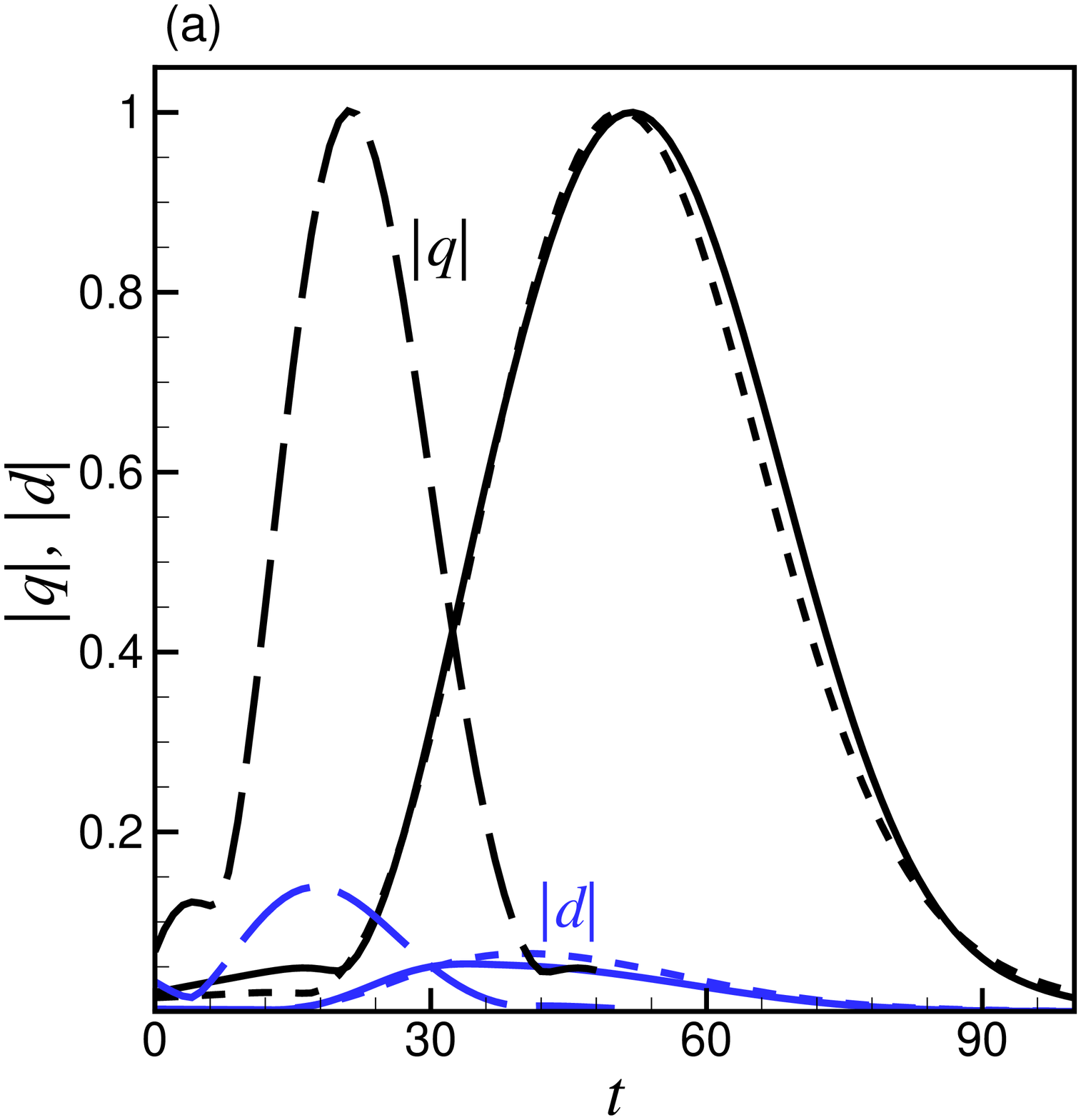}
           \includegraphics[scale=0.19]{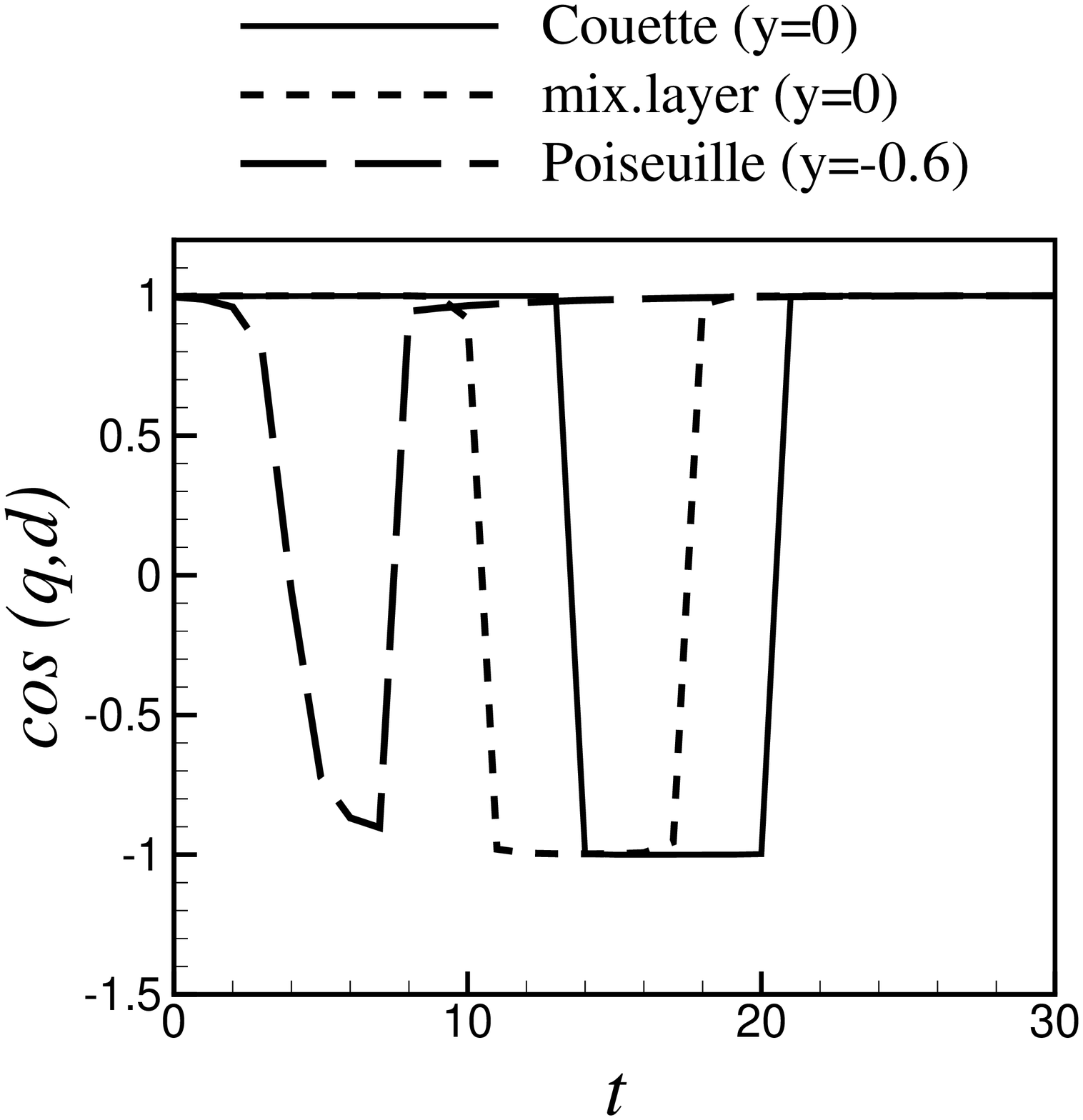}
           \includegraphics[scale=0.19]{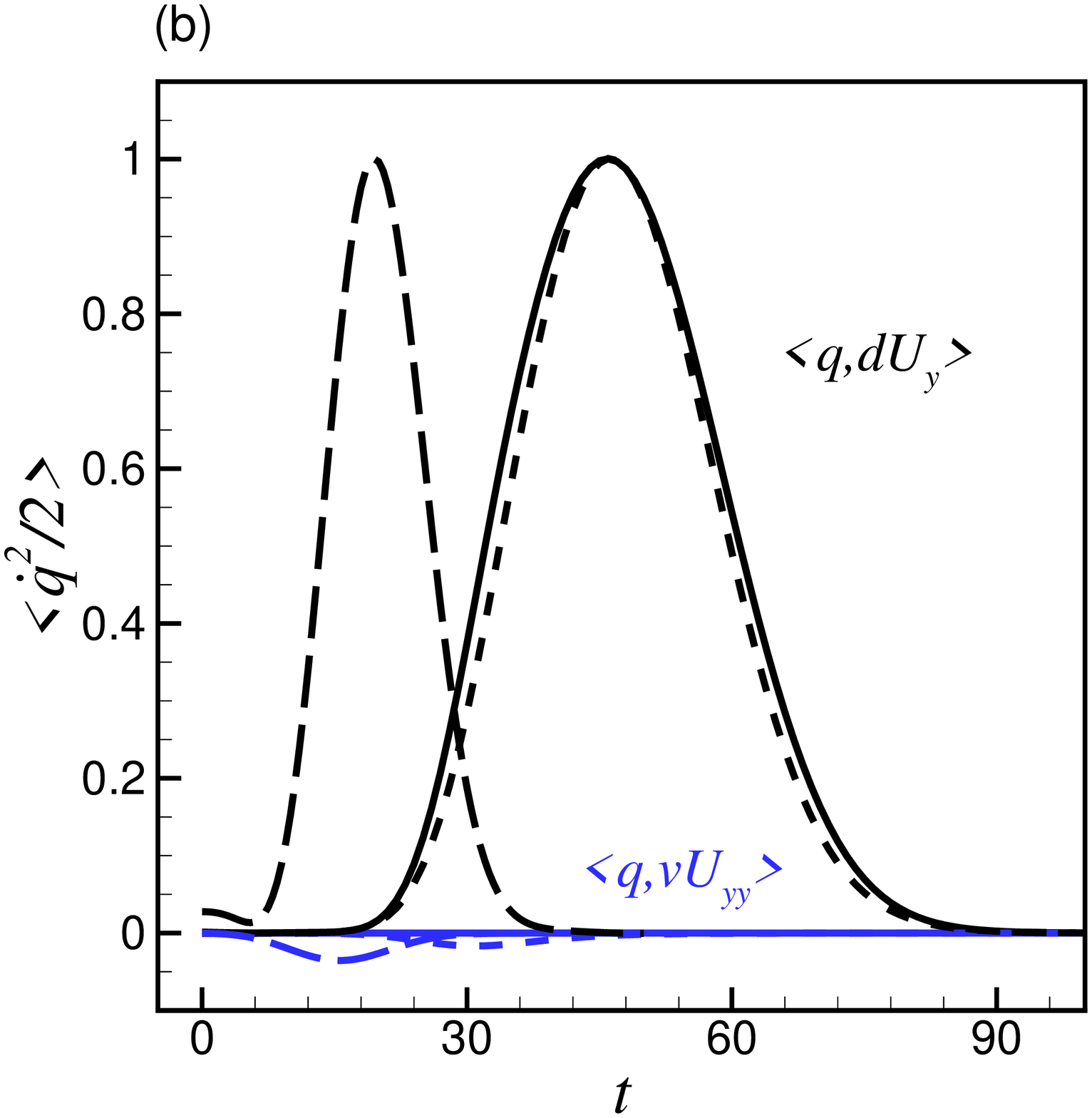}}
\caption{3D Growth of the planar divergent ($d$) and the span-wise vorticity ($q$) terms for the cases presented in Fig. 1,  for Couette (solid),Poiseuille (solid-dashed) and mixing layer (dashed) profiles. (a) Evolution of the absolute values of the two fields. Note that the amplification of $q$ is much larger than of $d$. (b) The normalized inner product between $d$ and $q$ at three selected levels (far enough from the boundaries) where the shear is positive and large. For the three profiles the fields are initially in phase, then become abruptly anti-phased, but return to be in phase at the stage of maximal growth. (c) Comparison between the two sources for growth of $q$. At the stage of maximal growth the vortex stretching term is larger by at least one order of magnitude than the vorticity advection term. Only the former is included in the minimal model.}
\label{fig:figure9}
\end{figure}

We now wish to focus on how the interplay between $d$ and $q$ affects the energy growth. Toward this end we invert $d$ and $q$ numerically to compute the stream-functions $\phi= \nabla^{-2}_2 d$, and $\psi = \nabla^{-2}_2 q$ in order to derive the divergent and vorticity induced velocity fields according to (3.6). The contributions of the four terms of equation (3.10) to the energy growth are then computed and presented in Fig. 5.

A similar qualitative behavior is identified for the three profiles. At the beginning, when the growth is mainly 2D, the Orr mechanism $-\<u_q v_q\>$, dominates. Later on, when $d$ amplifies $q$, the Orr mechanism increases accordingly but contributes toward a decay of the energy since the plane-wave like structures are tilted with the shear. Hence for 3D growth, the Orr mechanism is mostly an energy sink rather than a source\footnote{For the 2D bounded Couette flow the amplitude of $q$ remains constant (apart from small dissipation), and therefore energy growth and decay by the Orr mechanism should be almost symmetric. In Fig. 5. we also show the evolution of $-\<u_q v_q\>$ for the 2D case presented in Fig. 1. It indeed maximizes when the eddies are tilted against the shear in an angle of $\pi/4$, vanishes when they become perpendicular to the shear and finally becomes negative, in an anti-symmetric fashion, when titled with the shear.}. At this stage the pure divergent contribution $-\<u_d v_d\>$, is positive as expected, however this contribution is relatively small since $d$ itself remains small. It is clear from Fig. 5, and from the eddy geometry (Fig. 3), that the mixed term $-\<u_q v_d\>$, should be positive and large when the eddies are titled strongly with the shear (the more $-l$ becomes large). This mixed term is responsible for most of the 3D optimal energy growth and is able to overwhelm the large negative contribution of the Orr mechanism. The last combined term $-\<u_d v_q\>$, is negative but relatively small since by then $k<<|l|$.

\begin{figure}
\centering{\includegraphics[scale=0.18]{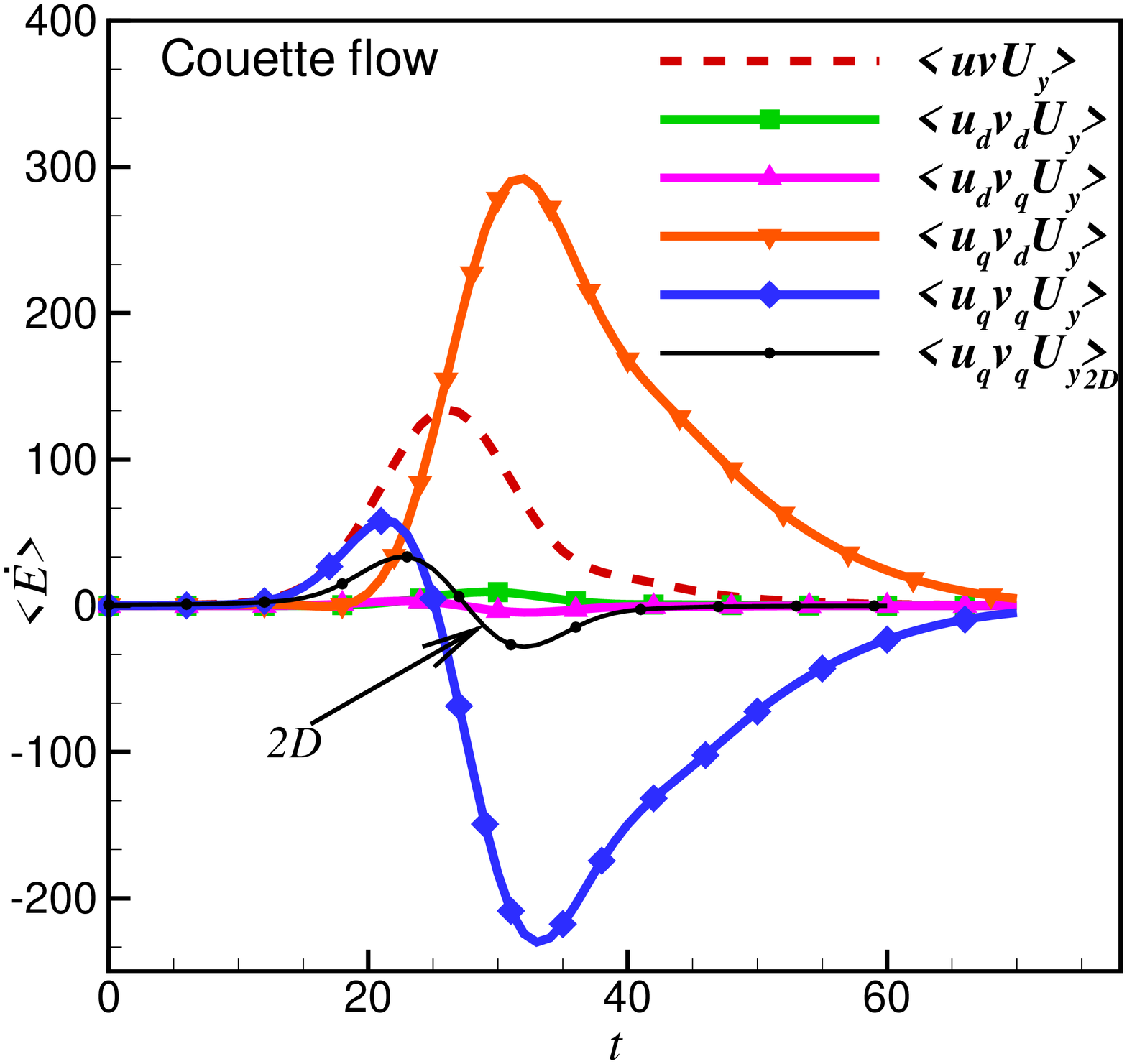}
           \includegraphics[scale=0.20]{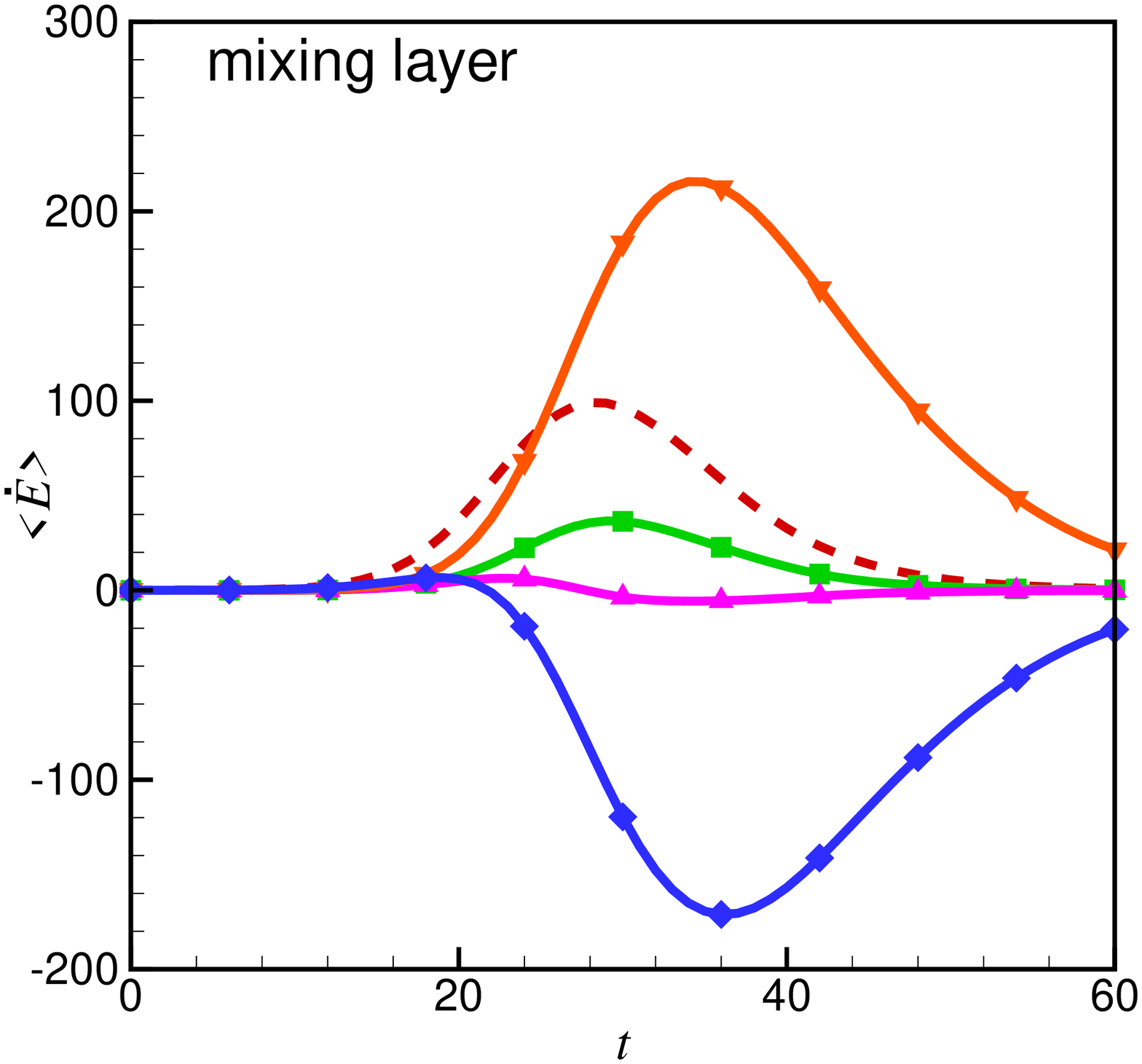}
           \includegraphics[scale=0.18]{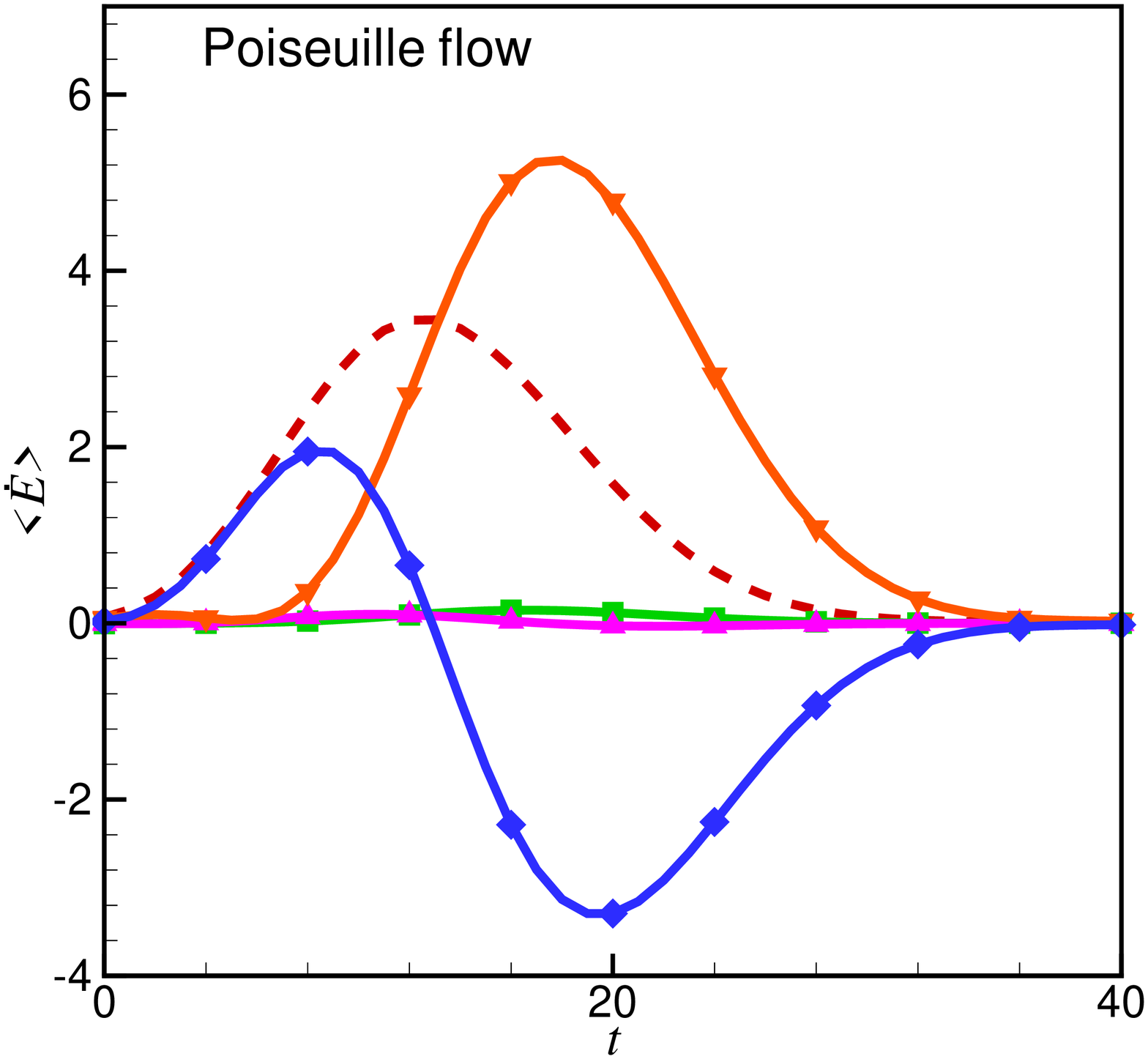}}
\caption{The various components of the Reynolds stress for the three 3D examples presented in Fig. 1. The behavior is robust: at the beginning of the evolution, when the eddies are tilted against the shear, the growth is dominated by the Orr term, $-\<u_q v_q\>$. When the eddies are turned to be tilted with the shear the Orr mechanism becomes the major energy sink (for the 2D dynamics in the Couette profile the Orr term is the only source for the Reynolds stress and it is presented for comparison in the left panel.)
The pure divergent contribution, $-\<u_d v_d\>$, turns from being negative to positive but remains small, while the mixed divergent rotational term, $-\<u_d v_q\>$, changes sign from positive to negative but remains small as well. It is the other mixed term, $-\<u_q v_d\>$, that becomes positive and large as the eddies become strongly titled with the shear. Indeed the latter term is responsible for the large 3D optimal growth.}
\label{fig:figure10}
\end{figure}

\section{Conclusions}

The energy growth mechanism which results from the Reynolds stress, $<-uv\der{\overline{U}}{y}>$, involves only the 2D shear plane dynamics, and is the same whether or not the perturbation itself is 2D or 3D. Nonetheless, the behavior of the optimal energy growth in 2D is very different from the one in 3D. This motivates us to reexamine the problem and seek for an  understanding which involves only the dynamics on the shear plane. The common explanation for the 3D growth via the lift-up mechanism, involves the span-wise variation of the cross-stream velocity and the generation of cross-stream vorticity. Since in planar 2D dynamics there is no span-wise variation and the cross-stream vorticity is zero by definition, it is difficult to compare between the 2D and the 3D optimal dynamics from the lift-up perspective.

For incompressible flow the 2D dynamics results only from the rotational component (the span-wise vorticity) of the flow, whereas in 3D it results both from the rotational and the divergent components on the shear plane. A mutual growth of these two components could explain why the 3D optimal growth is much larger than the 2D one. Numerical computations of the optimal dynamics for three different canonical profiles of plane parallel shear flows (Couette, Poiseuille, and mixing layer) reveal a generic behavior of the perturbations at the stage of maximal growth. The perturbations are then organized as localized plane-wave like structures, in regions where the background shear is the largest. The waves are tilted with the shear, and the planar vorticity and divergence fields are in (out of) phase when the background shear is positive (negative). This picture is very different from the optimal growth of 2D perturbations by which energy grows when eddies are tilted against the shear.

The insensitivity of the optimal dynamics to the shear curvature and the robustness of the plane-wave like structure allow the construction of a minimal model, of a plane-wave deformed by a constant shear, to explore the essence of the 3D energy growth from the divergent-rotational interplay perspective. The divergent field affects the rotational one by vortex stretching in a straightforward manner. Indeed, as obtained by the numerical simulations for background positive shear, span-wise vorticity grows when it is in phase with the divergent field. The growth mechanism of the planar divergent field is less obvious. It occurs when the pressure perturbation is negative and results both from a differential advection of background momentum by the perturbation field, and from a differential advection of perturbation momentum by the background flow. When the divergent and rotational components are in phase the contribution of the rotational field to the divergent growth is always negative. The contribution of the divergent field to its own growth is positive only when the eddies are tilted with the shear. Hence, in order to obtain a mutual growth of the two fields they should be in phase and tilted with the shear. Then a large growth in the span-wise vorticity field is accompanied by a modest growth in the divergence field.

A mutual growth of the divergent and rotational components is not enough by itself to ensure a large energy growth when eddies are tilted with the shear. Decomposing the Reynolds stress in terms of the various contributions from the divergent and the rotational fields indicates that the mixed term $<-u_qv_d\der{\overline{U}}{y}>$, which can be very large and positive when the plane-wave like structures are strongly tilted with the shear, is mostly responsible for the large 3D energy growth. The other terms also contribute - the rotational component $<-u_q v_q\der{\overline{U}}{y}>$ generates a large negative energy growth via the Orr mechanism, the divergent term $<-u_d v_d\der{\overline{U}}{y}>$ acts as a modest source for growth, and the combined divergent-rotational contribution, $<-u_d v_q\der{\overline{U}}{y}>$ is negative.

The linear analysis of optimal dynamics presented here may also be relevant for nonlinear flows, for which the Reynolds stress is also the instantaneous source for energy growth. We have shown that the vortex stretching mechanism which acts only in 3D, leads to the fundamentally different energy growth dynamics in 3D compared to 2D. Vortex stretching also leads to the basic difference in the direction of energy cascade in 2D and 3D flows. One might argue then that in 2D, the Orr mechanism contributes to the inverse cascade since the maximum energy is attained when the cross-stream wavenumber vanishes, so that the total wavenumber decreases. In contrast, in 3D, the vortex stretching dynamics leads to amplification of energy when the cross-stream wavenumber is large, in line with the direct cascade mechanism.

\appendix
\section{}\label{app}
{\noindent \Large \bf Analytical solution of sheared plane-wave in unbounded Couette flow}

\setcounter{equation}{0}

Define  ${\bf K}^2_V \equiv {(k^2+m^2)}$, the solution for equation (3.8) is:
\be
\hat q (t)= \hat q_0 + \hat d_0\Lambda t - {m^2 \hat p_0\over 2 k^2 \Lambda}{{\bf K}_0^4\over {\bf K}_V^2 }{\left(\frac{\Lambda k l_0}{{\bf K}_0^2}t
+{\frac{l}{{|\bf K}_V|} \left[\arctan \left(\l\over{|\bf K}_V|\right)-\arctan \left(\l_0\over{|\bf K}_V|\right)\right] }\right)}
\ee
\be
\hat d (t)= \hat d_0 +{m^2 \hat p_0\over 2 k \Lambda}{{\bf K}_0^4\over {\bf K}_V^2 }{\left(\frac{l}{{\bf K}^2}
- \frac{l_0}{{\bf K_0}^2}+{\frac{1}{{|\bf K}_V|} \left[\arctan \left(\l\over{|\bf K}_V|\right)-\arctan \left(\l_0\over{|\bf K}_v|\right)\right] }\right)}
\ee
where $\hat p = \({|{\bf K}_0|\over |{\bf K}|}\)^4 \hat p_0$, and $\hat v = \({|{\bf K}_0|\over |{\bf K}|}\)^2 \hat v_0$.

As an example, the different components of the energy growth are presented in Fig. 6 for the plane wave solution. The background shear and the stream-wise and span-wise wavenumbers are the same as in Fig. 1. for the 3D bounded Couette flow example. The initial cross-stream wavenumber $l_0 = k\Lambda t_{perp}=9$, is chosen so that $l=0$ corresponds to the normalized time $\Lambda t_{perp} = 18$, in which the optimal perturbation span-wise vorticity contours become perpendicular to the shear. Comparison between Fig. 5a. and Fig. 6 reveals that as expected, the plane-wave does not mimic the energy growth when it is tilted against the shear. Nonetheless, at the stage of maximum growth the essence of the behavior is similar, where the mixed $d$--$q$ term,  $<-u_qv_d\Lambda>$, overwhelms the negative contribution of the Orr mechanism $<-u_qv_q\Lambda>$.

\begin{figure}
\centerline{\includegraphics[scale=0.25]{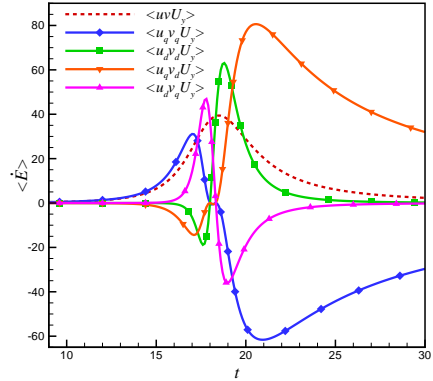}}
\caption{Same as in Fig. 5 but for the plane-wave solution of (A.1,2). The initial perturbation is normalized by the energy norm.
The stream and the span-wise wavenumbers $(k,m)=(0.5,1)$, are as in the example for the Couette flow in Fig. 1. The initial cross-stream wavenumber $l_0 = k\Lambda t_{perp} =9$, where $\Lambda t_{perp}$ is the normalized time by which the eddy stream-wise vorticity contours become untilted in the bounded Couette simulation, presented in Fig. 1. In the stage of maximal growth Figs. 5 and 6 present a similar qualitative behavior.}
\label{fig:A1}
\end{figure}

\bibliographystyle{jfm}

\bibliography{jfm-Lena}

\end{document}